\documentclass[sort&compress
    ,final            
  ]
  {aipproc}

\layoutstyle{6x9}

\begin{document}

\title{Deconfinement to Quark Matter in Neutron Stars - The Influence of Strong Magnetic Fields}


\author{V. Dexheimer}{
  address={UFSC, Florianopolis, Brazil}
  ,altaddress={Gettysburg College, Gettysburg, PA, USA} 
}

\author{R. Negreiros}{
  address={UFF, Niter\'oi, Brazil}
   ,altaddress={FIAS - Johann Wolfgang Goethe University, Frankfurt, DE}
}

\author{S. Schramm}{
  address={FIAS - Johann Wolfgang Goethe University, Frankfurt, DE}
}

\author{M. Hempel}{
  address={University of Basel, Basel, CH}
}

\begin{abstract}
We use an extended version of the hadronic SU(3) non-linear realization of the sigma model that also includes quarks to study hybrid stars. Within this approach, the degrees of freedom change naturally as the temperature/density increases. Different prescriptions of charge neutrality, local and global, are tested and the influence of strong magnetic fields and the anomalous magnetic moment on the particle population is discussed.  
\end{abstract}

\maketitle


\section{Introduction and Model Description}

In order to describe hybrid stars we make use of a model that contains both hadronic and quark degrees of freedom. At high densities and/or temperatures the hadrons are replaced by quarks, as their effective masses increase (hadrons) and decrease (quarks). The aforementioned model (see \cite{Dexheimer:2009hi} for details) is an extended version of the SU(3) non-linear realization of the sigma model \cite{Papazoglou:1998vr,Dexheimer:2008ax,Negreiros:2010hk} within the mean-field approximation. Changes in the order parameters $\sigma$ and $\Phi$ signal chiral symmetry restoration and quark deconfinement, respectively. The potential for $\Phi$ is an extension of the Polyakov loop potential \cite{Roessner:2006xn} modified to also depend on baryon chemical potential. In this way the model is able to describe the entire QCD phase diagram and agrees with lattice QCD constraints \cite{Fodor:2004nz}. The phase transitions at low temperatures and high densities are of first order, while at high temperatures and low densities the model exhibits a smooth crossover, in accordance with lattice QCD
results.. 

\begin{figure}
  \includegraphics[height=.3\textheight]{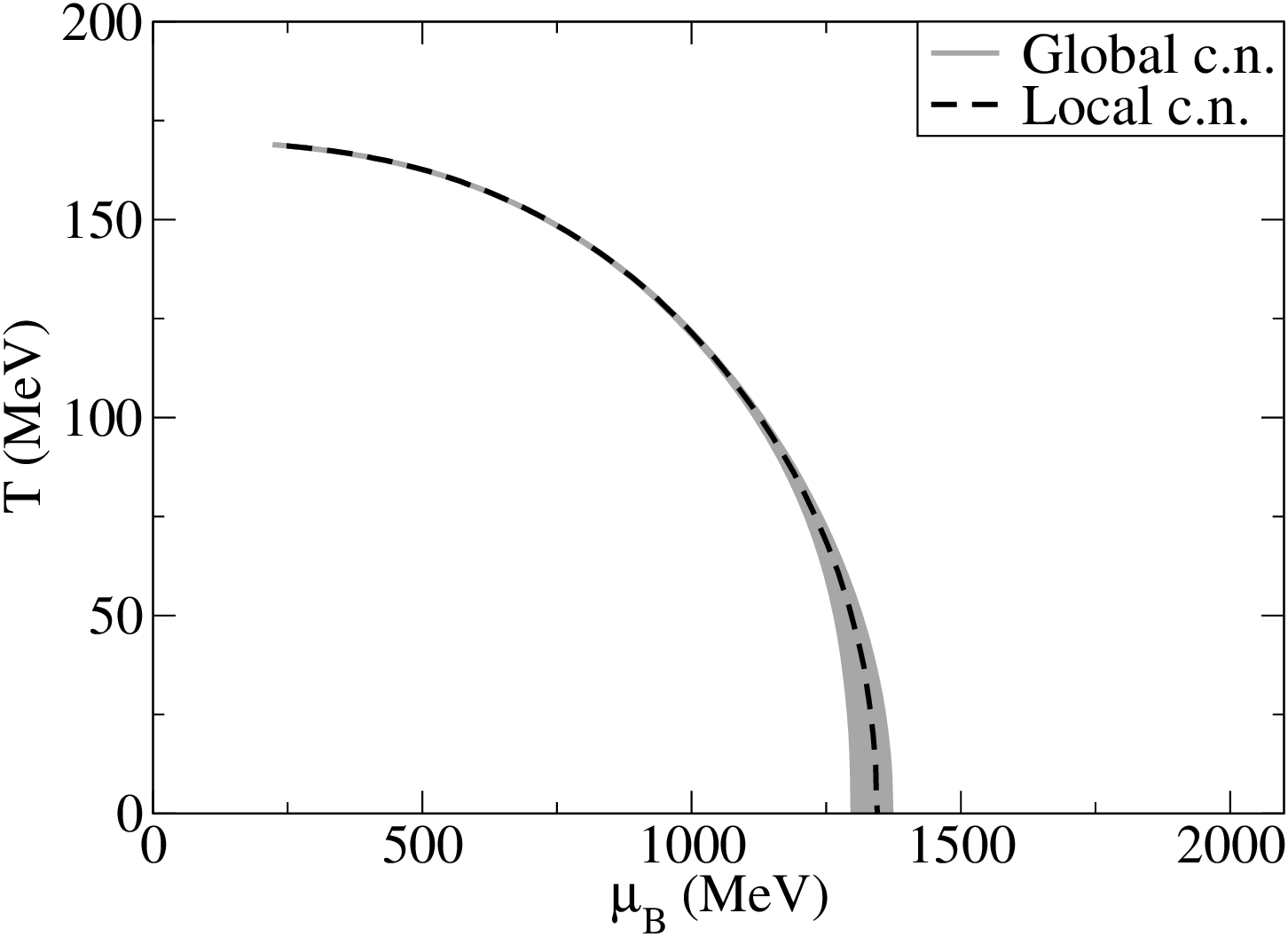}
  \caption{Temperature versus baryon chemical potential for neutron star matter assuming local and global charge neutrality.}
  \label{plot1}
\end{figure}

\begin{figure}
  \includegraphics[height=.3\textheight,clip,trim=0 0 0 1]{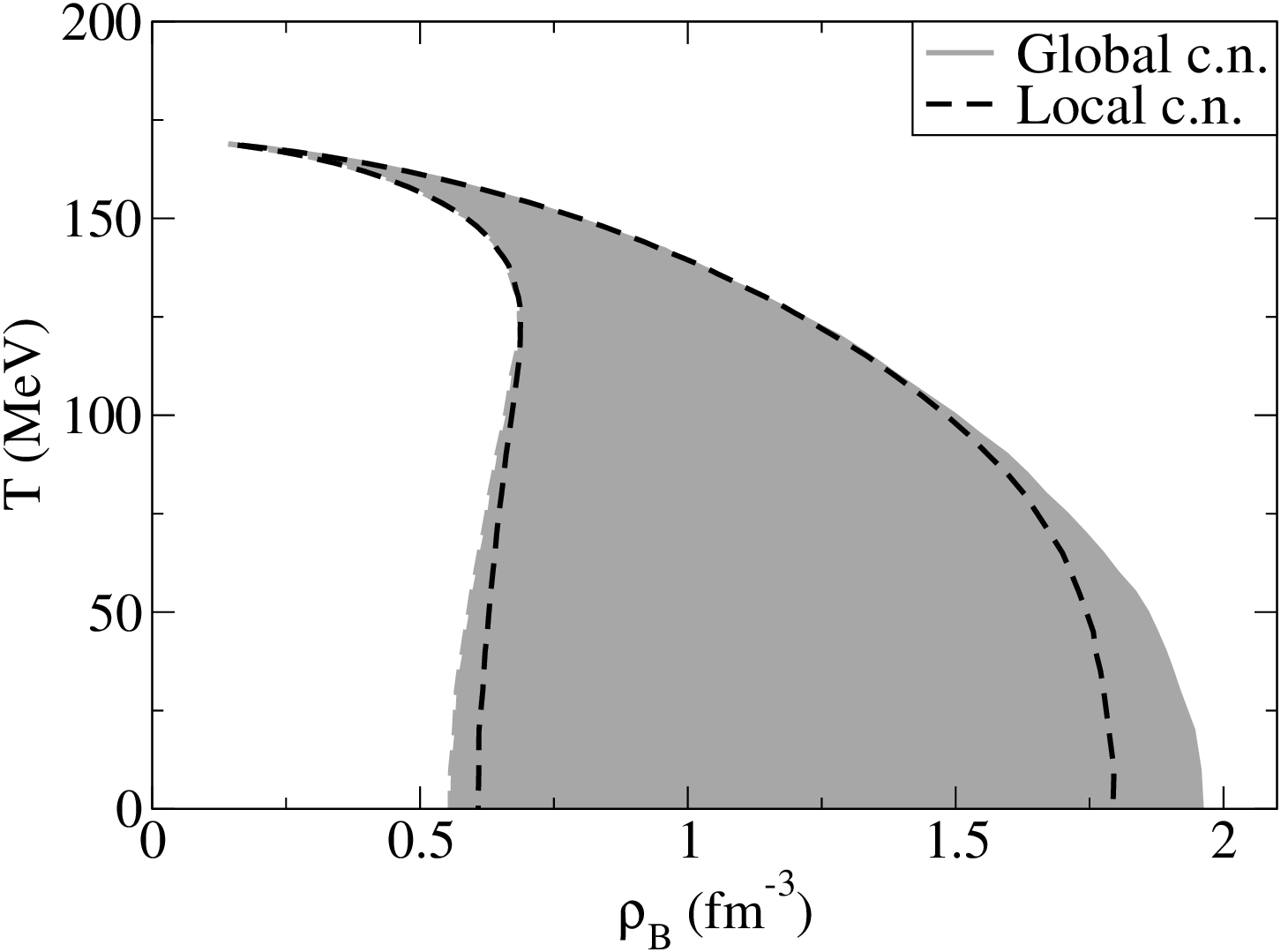}
  \caption{Temperature versus baryon density for neutron star matter assuming local and global charge neutrality.}
  \label{plot2}
\end{figure}

For neutron star calculations, matter is further required to be charge neutral and in chemical equilibrium. In the case of a first-order phase transition, we can either require each phase to be individually charge neutral (local charge neutrality) or allow both phases to be charge neutral only when combined (global charge neutrality). Fig.~\ref{plot1} shows how the resulting QCD phase diagram looks like following both of these prescriptions. When global charge neutrality is assumed, a mixed phase appears. It extends over a range of baryon chemical potentials, that corresponds to a range of baryon densities (Fig.~\ref{plot2}). If local charge neutrality is assumed instead, the phase
transition occurs for a given temperature at a single value of the chemical potential. Nevertheless, a small step in baryon chemical potential at the phase transition corresponds to a large jump in baryon density. Note that for local charge neutrality all densities within these limits correspond to the same pressure and, therefore, cannot occupy a physical space in a star.

\begin{figure}
  \includegraphics[height=.3\textheight]{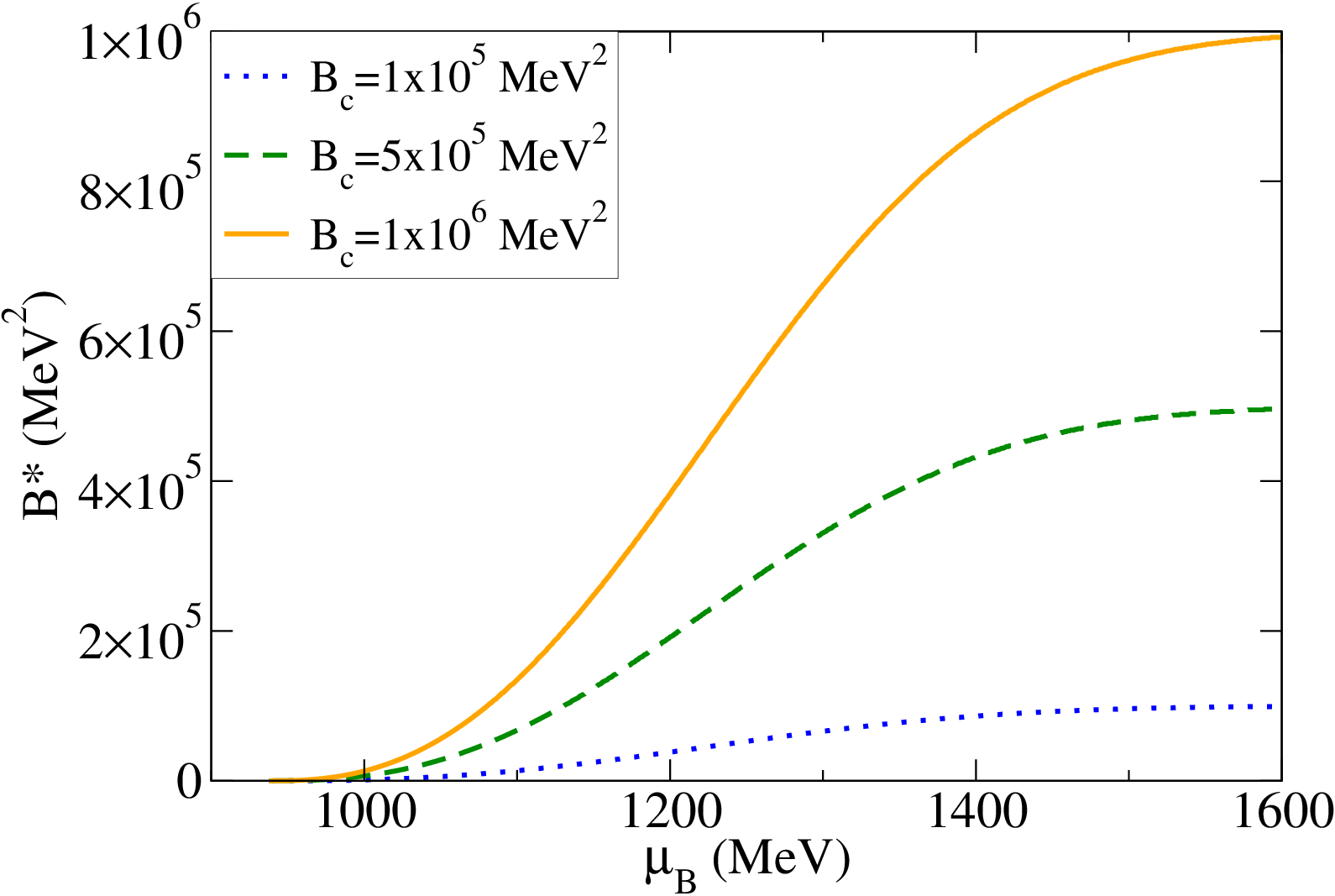}
  \caption{Effective magnetic field as a function of baryonic chemical potential shown for different central magnetic fields. Recall that using Gaussian natural units $1$ MeV$^2=1.44\times10^{13}$ G. \label{Beff}}
\end{figure}

The next step towards a complete description of neutron stars is the inclusion of the effect of magnetic fields in the model. The highest magnetic field observed on the surface of magnetars is on the order of $10^{15}$ G. The highest possible magnetic field in the center of stars, on the other hand, can only be estimated through models, even when applying the Virial theorem. Some results indicate limiting magnetic fields ranging between $B=10^{18}-10^{20}$ G ~\cite{Bocquet:1995je,Cardall:2000bs,1991ApJ...383..745L,Chakrabarty:1997ef,Bandyopadhyay:1997kh,Broderick:2001qw,Ferrer:2010wz,
Malheiro:2004sb}.
In order to avoid hydrodynamical instabilities due to pressure anysotropy \cite{Chaichian:1999gd,PerezMartinez:2005av,PerezMartinez:2007kw,Huang:2009ue,Paulucci:2010uj} and still be able to include strong magnetic fields, we consider in our calculation a magnetic field in the z-direction that varies with baryon chemical potential. More precisely the assumed magnetic field $B^*$ runs from a surface value $B_{surf}=69.25$ MeV$^2=10^{15}$ G (when $\mu_B=938$ MeV) to different central values $B_c$ at extremely high baryonic chemical potential following \cite{Dexheimer:2011pz}
\begin{equation}
B^*(\mu_B)=B_{surf}+B_c[1-e^{b\frac{(\mu_B-938)^a}{938}}],
\end{equation}
with $a=2.5$, $b=-4.08\times10^{-4}$ and $\mu_B$ given in MeV. As can be seen in Fig.~\ref{Beff}, the values of effective magnetic field only approach $B_c$ at very high baryonic chemical potentials and, in practice, only about $70\%$ of $B_c$ can be reached inside neutron stars. The use of baryon chemical potential instead of density was chosen to prevent discontinuities in the magnetic field at the phase transition where, as we have already mentioned, there is a jump in baryon density. Nevertheless, the constants $a$ and $b$ and the form of the $B^*$ expression were chosen to reproduce (in the absence of quarks) the effective magnetic field curve as a function of density from Refs.~\cite{Bandyopadhyay:1997kh,Mao:2001fq,Rabhi:2009ih}.

\section{Results and Conclusions}

The magnetic field in the z-direction forces the eigenstates in the x and y directions of the charged particles to be quantized into Landau levels. The energy levels of all baryons are further split due to the alignment/anti-alignment of their spin with the magnetic field (anomalous magnetic moment effect AMM). A complete analysis of these effects together with chiral symmetry restoration and quark deconfinement within the extended version of the SU(3) non-linear realization of the sigma model on neutron star observables can be found in \cite{Dexheimer:2011pz}. Similar studies using different models can be found in Refs.~\cite{Menezes:2009uc,Menezes:2008qt,2011PhRvC..83f5805A}.

\begin{figure}
  \includegraphics[height=.3\textheight]{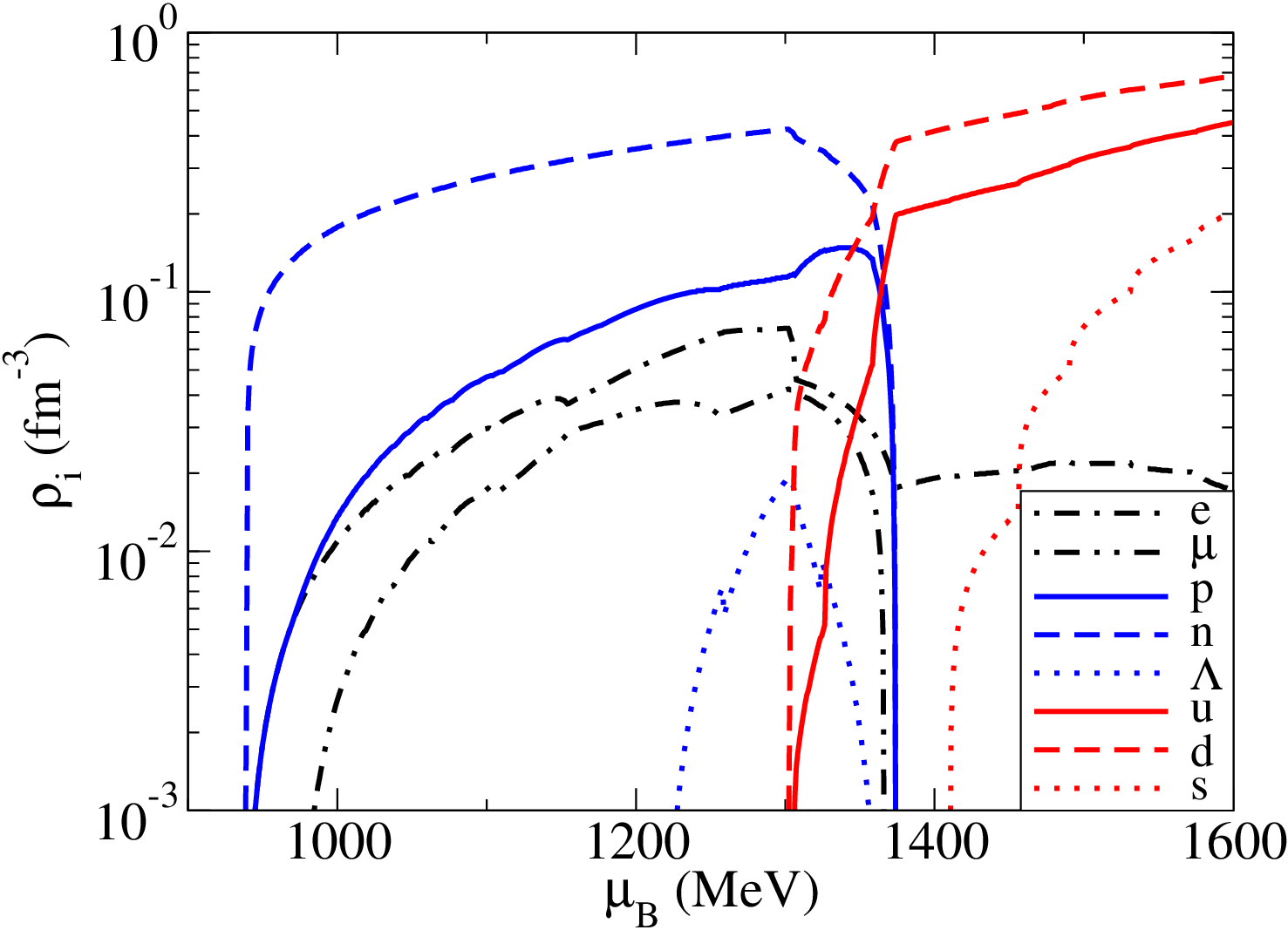}
  \caption{Particle densities as a function of baryonic chemical potential assuming global charge neutrality for $B_c=5\times10^5$ MeV$^2=7.22\times10^{18}$ G including AMM. \label{popB}}
\end{figure}

In this work, we chose to focus on the analysis of magnetic field effects on the chemical composition of the neutron star, using global charge neutrality. More specifically, Fig.~\ref{popB} shows the density of fermions (with quark number densities divided by 3) when a central magnetic field $B_c=5\times10^5$ MeV$^2=7.22\times10^{18}$ G with AMM is considered. In the mixed phase the hadrons disappear smoothly as the quarks appear. The hyperons, despite being included in the calculation, are suppressed by the appearance of quarks and only a very small amount of $\Lambda$'s appear. The strange quarks appear after the other quarks and do not make substantial changes in the system. The "wiggles" in the charged particle densities mark when their Fermi energies cross the discrete threshold of a Landau level. The charged particle population is enhanced due to $B$, as their chemical potentials increase.

\begin{figure}
  \includegraphics[height=.3\textheight]{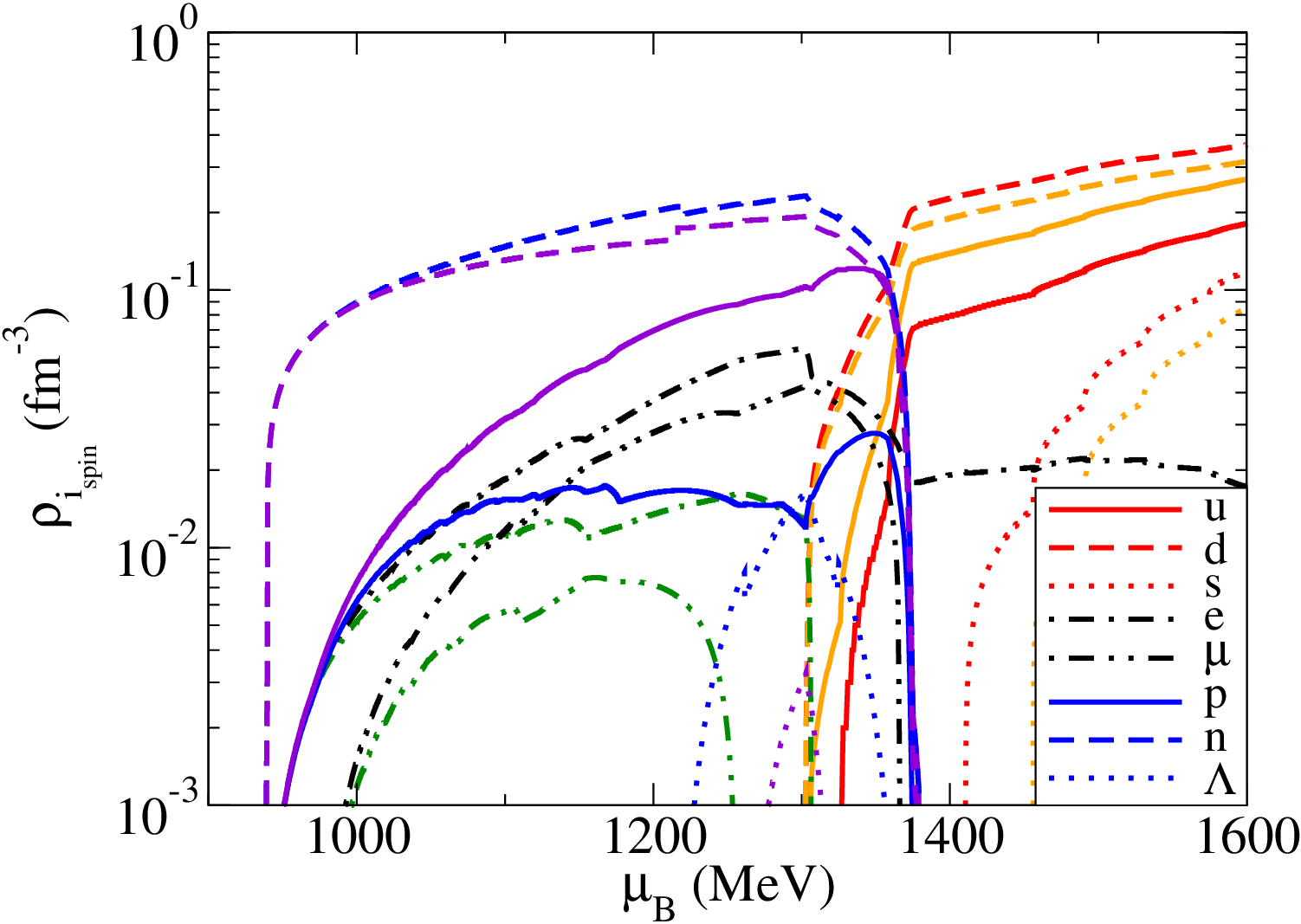}
  \caption{Same as Fig.~\ref{popB} but differentiating particles by spins. Black, blue and red stand for negative spin projections, while green, purple and orange stand for positive spin projections, respectively.\label{popspin}}
\end{figure}

Although the AMM is known to make the EOS stiffer, it does not have a very significant effect in the particle population \cite{Broderick:2001qw}. This fact can be easily understood in terms of polarization, when, instead of looking at the total particle density (sum of spin up and down particle densities) for each species, we look at the spin up/spin down particle densities separately. Fig.~\ref{popspin} shows that some of these particles are enhanced while others are suppressed. 
For non-charged baryons, the ones with positive spin projection are suppressed due to the increase in their effective masses caused by the anomalous magnetic moment ${m_i}^* \to {m_i}^* - s \kappa_i B$ combined with negative $\kappa_i$ (like neutrons and $\Lambda$'s). The baryons with positive spin projection and positive $\kappa_i$ (like the protons) are enhanced by the AMM. Particles with negative spin projection have the exactly opposite behavior.

For particles that are not strongly affected by the AMM, spin polarization can be understood by looking at the definition of Landau level $\nu=l+\frac{1}{2}-\frac{s}{2} \frac{{Q_e}_i}{|{Q_e}_i|}$. Particles with spin projection $s$ opposite to their charge sign ${Q_e}_i/{|{Q_e}_i|}$ cannot have $\nu=0$, even when their orbital angular momentum $l$ is equal to zero, causing them to have a smaller density when summing over all levels. Particles with spin projection equal to their charge sign are enhanced. Note that the electrons in the quark phase are fully polarized.

We have shown in this work some possible effects of strong magnetic fields in hybrid stars. The presence of different hadronic and quark degrees of freedom makes the extended version of the SU(3) non-linear realization of the sigma model an ideal tool for such an analysis, in all of the different possible phases. In the future, we intend to calculate the influence of strong magnetic fields in chiral symmetry restoration and quark deconfinement for neutron star matter at finite temperatures. Such kind of analysis can help us have a better understanding of the QCD phase diagram. 


\begin{theacknowledgments}

V. D. acknowledges support from CNPq (National Counsel of Technological and Scientific Development - Brazil).
M. H. acknowledges support from the High Performance and High Productivity Computing (HP2C) project, the Swiss National Science Foundation (SNF) and ENSAR/THEXO. M. H. is also grateful for support from CompStar, a research networking program of the ESF.

\end{theacknowledgments}



\bibliographystyle{aipproc}   

\bibliography{template-6s}

\IfFileExists{\jobname.bbl}{}
 {\typeout{}
  \typeout{******************************************}
  \typeout{** Please run "bibtex \jobname" to optain}
  \typeout{** the bibliography and then re-run LaTeX}
  \typeout{** twice to fix the references!}
  \typeout{******************************************}
  \typeout{}
 }

\end{document}